\documentclass[superscriptaddress,amsmath,amssymb,aps,prl]{revtex4-2}

\usepackage{graphicx}
\usepackage{dcolumn}
\usepackage{bm}
\usepackage[colorlinks]{hyperref}

\newcommand{\uck}[1]{\o}

\newcommand{\ket}[1]{\mbox{$|#1\protect\rangle$}}

\newcommand{\bra}[1]{\mbox{$\protect\langle#1|$}}

\newcommand{\tr}{\text{tr}}

\begin{document}

\title{Quantum logical entropy: fundamentals and general properties}

\author{B. Tamir}
\thanks{These two authors contributed equally to this work.}
\affiliation{Faculty of Interdisciplinary Studies, Bar-Ilan University, Ramat Gan 5290002, Israel}

\author{I. L. Paiva}
\thanks{These two authors contributed equally to this work.}
\affiliation{Faculty of Engineering and the Institute of Nanotechnology and Advanced Materials, Bar-Ilan University, Ramat Gan 5290002, Israel}

\author{Z. Schwartzman-Nowik}
\affiliation{School of Computer Science and Engineering, The Hebrew University, Jerusalem 91904, Israel}
\affiliation{Faculty of Engineering and the Institute of Nanotechnology and Advanced Materials, Bar-Ilan University, Ramat Gan 5290002, Israel}

\author{E. Cohen}
\affiliation{Faculty of Engineering and the Institute of Nanotechnology and Advanced Materials, Bar-Ilan University, Ramat Gan 5290002, Israel}

\begin{abstract}
Logical entropy gives a measure, in the sense of measure theory, of the distinctions of a given partition of a set, an idea that can be naturally generalized to classical probability distributions. Here, we analyze how fundamental concepts of this entropy and other related definitions can be applied to the study of quantum systems, leading to the introduction of the quantum logical entropy. Moreover, we prove several properties of this entropy for generic density matrices that may be relevant to various areas of quantum mechanics and quantum information. Furthermore, we extend the notion of quantum logical entropy to post-selected systems.
\end{abstract}

\maketitle

\section{Introduction}

Entropy is one of the paramount concepts in probability theory and physics. Even though information does not seem to have a precise definition, the Shannon entropy is seen as an important measure of information about a system of interest, while the Gibbs entropy plays a similar role in statistical mechanics. A possible and, in a sense, natural extension of these classical measures to the quantum realm is the von Neumann entropy. Despite playing a fundamental role in many applications in quantum information, the von Neumann entropy was criticized on several different grounds \cite{brukner1999operationally, brukner2001conceptual, giraldi2001quantum}. In a nutshell, while classical entropy indicates one's ignorance about the system \cite{cover2006elements}, quantum entropy seems to have a fundamentally different flavor, corresponding to an \textit{a priori} inaccessibility of information or the existence of non-local correlations. In this perspective, classical entropy concerns subjective/epistemic indefiniteness, while quantum entropy is associated with some form of objective/ontological indefiniteness \cite{ellerman2013information}, although this reasoning can be contested. To address conceptual problems like this one, the non-additive Tsallis entropy and other measures were proposed \cite{tsallis1988possible, brukner1999operationally, manfredi2000entropy}.

\begin{figure}
    \centering
    \includegraphics[width=.5\columnwidth]{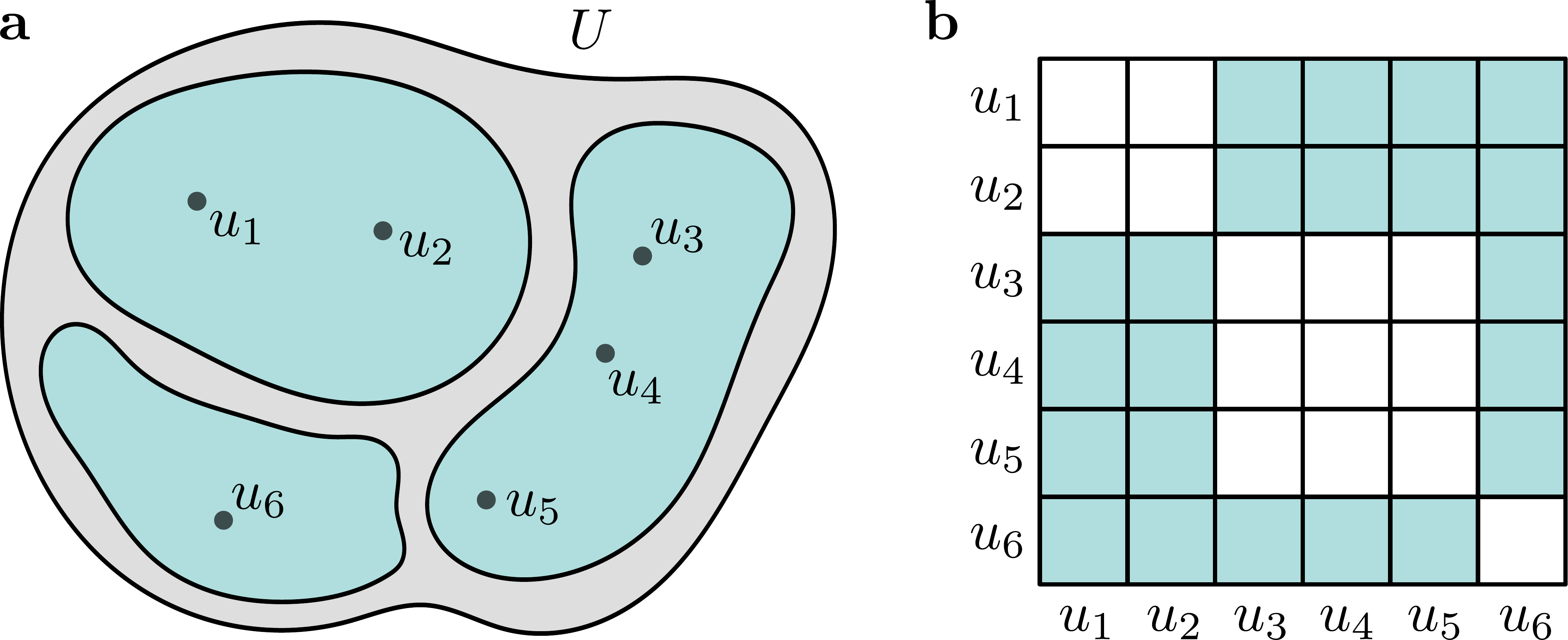}
    \caption{\textbf{Pictorial representation of a set with a partition and its distinction.} \textbf{(a)} A set $U$ with six elements $u_i$ is divided in a partition $\pi$ with three blocks (green areas). \textbf{(b)} The set $U\times U$ has each of its points characterized by a square in a 2D mesh. Points that belong to $\text{dit}(\pi)$, i.e., pairs whose components belong to different blocks are colored in green.}
    \label{fig:example}
\end{figure}

Classical logical entropy was recently introduced by Ellerman \cite{ellerman2013introduction, ellerman2014introd} as an informational measure arising from the logic of partitions. As such, this entropy gives the distinction of partitions of a set $U$. A partition $\pi$ is defined as a set of disjoint parts of a set, as exemplified in Figure \ref{fig:example}a. The set can be thought of as being originally fully distinct, while each partition collects together blocks whose distinctions are factored out. Each block represents elements that are associated with an equivalence relation on the set. Then, the elements of a block are indistinct among themselves while different blocks are distinct from each other, given an equivalence relation.

With these concepts in mind, it seems that the extension of this framework of partitions and distinctions to the study of quantum systems may bring new insights into problems of quantum state discrimination, quantum cryptography, and quantum channel capacity. In fact, in these problems, we are, in one way or another, interested in a distance measure between distinguishable states, which is exactly the kind of knowledge the logical entropy is associated with.

This work is an updated and much extended version of a previous preprint \cite{tamir2014logical}. In this new version, like in the original one, we focus mostly on basic definitions and properties of the quantum logical entropy. Other advanced topics were either treated in previous research, like in \cite{tamir2015holevo}, or are left for future investigation. However, as it will be further elaborated upon across the text, the results presented here lay the groundwork for various theoretical applications --- even for scenarios involving post-selected systems.

To set the framework, we start with a brief overview of classical logical entropy. For that, consider a finite set $U$ and a partition $\pi = \{B_i\}$ of $U$, where each block $B_i$ is a disjoint part of $U$. Moreover, denote by $\text{dit}(\pi)$ the distinction of the partition $\pi$, i.e., the set of all pairs $(u,u')\in U\times U$ such that $u$ and $u'$ are not in the same block $B_i$ of the partition $\pi$, as illustrated in Figure \ref{fig:example}b.

The logical entropy $L_\pi(U)$ is defined as
\begin{equation}
    L_\pi(U) = \frac{|\text{dit}(\pi)|}{|U\times U|},
\end{equation}
where $|\cdot|$ denotes the cardinality of the set. In other words, $L_\pi(U)$ is the probability of getting elements from two different blocks $B_i$ after uniformly drawing two elements from $U$. Therefore, it is a measure of average distinction.

If each element of $U$ is assumed to be equally probable, we can write $p_{B_i} = \frac{|B_i|}{|U|}$ and, thus,
\begin{equation}
    L_\pi(U) = 1- \sum_{B_i\in \pi} p_{B_i}^2.
    \label{eq:cle-part}
\end{equation}

Furthermore, if a probability $p_k$ is assigned to each element $u_k\in U$, the above expression can be applied to the partition $\pi = 1_U$ with element-blocks given by $\{u_i\}$. This leads to
\begin{equation}
    L(\{p_i\}) \equiv L_{1_U}(U) = 1- \sum_i p_i^2 = \sum_i p_i (1-p_i).
    \label{eq:classical-le}
\end{equation}
Therefore, $L(\{p_i\})$ is the probability to draw two different elements $u_i$ consecutively. To generalize it even further, one may consider an arbitrary set $U$ with a countable number of blocks. In this case, the idea of logical entropy is extended to countable probability distributions.

Starting from this definition, other concepts, like logical divergence, logical relative entropy, logical conditional entropy, and logical mutual information \cite{ellerman2013introduction}, can also be introduced along the same lines.

It should be noted that the formula for logical entropy is rooted in the history of information theory, preceding by at least a century its relation to the logic of partitions introduced by Ellerman. In fact, it can be traced back to Gini's index of mutability \cite{gini1912variabilita}. Also, Polish Enigma crypto-analysts and, later, Turing used the term ``repeat rate'' \cite{rejewski1981polish, patil1982diversity, good1982comment}, which is just the complement of the logical entropy. Moreover, it coincides with the \textit{Tsallis entropy of index 2} and, as a result, resembles the information measure suggested by Brukner and Zeilinger \cite{brukner1999operationally, brukner2001conceptual} and Manfredi \cite{manfredi2000entropy} for applications in quantum mechanics .

Here, we follow standard methods from quantum information (e.g., \cite{nielsen2000quantum}) to extend the notion of logical entropy to quantum states. In this approach, the set $U$ becomes the state of a quantum system, i.e., an element of a Hilbert space. With that, besides adding new results, we generalize the ones originally presented in \cite{ellerman2014partitions}, supporting them with formal proofs regarding quantum density matrices. We hope that the discussion we present will shed new light on this intriguing informational measure.

Since the writing of the original manuscript \cite{tamir2014logical}, additional works in this area have elaborated on the topic \cite{ellerman2016classicalquantum, ellerman2016classical, ellerman2017new, ellerman2017introduction, ellerman2018introduction, ellerman2018logical}. Still, we expect that the updates added here will provide new insights into the subject. In fact, within the next section, we present a new road to the introduction of quantum logical entropy that makes its connection with the idea of partitions present in the classical definition more evident. After that, in the section with the properties of quantum logical entropies, we prove some results that were not present in the original manuscript and simplify the proofs to others that were already on it whenever possible. Moreover, we added a section where new definitions of logical entropy for post-selected systems are introduced. To conclude this work, we further discussions some aspects of our study in the final section.

\section{Quantum logical entropy}

Every projection-valued measure (PVM) defines a partition $\pi$ in a Hilbert space in the sense that it provides a direct-sum decomposition of the space \cite{ellerman2016quantum, ellerman2018quantum}. In fact, recall that a PVM is characterized by parts $B_i=|b_i\rangle\langle b_i|$ such that $\sum_i B_i = I$. Then, the probability that two consecutive measurements of copies of a state $\rho$ in such a Hilbert space will belong to distinct parts is
\begin{equation}
    \begin{aligned}
        L_\pi(\rho) &\equiv \sum_i \tr(B_i\rho) [1 - \tr(B_i\rho)] \\
        &= \sum_i \tr(B_i\rho B_i) - \sum_i [\tr(B_i\rho B_i)]^2 \\
        &= \tr(\sum_i B_i\rho B_i) - \tr[\sum_i (B_i\rho B_i)^2] \\
        &= \tr\rho' - \tr(\rho')^2 \\
        &= \tr[\rho' (I-\rho')],
    \end{aligned}
    \label{eq:qle-part}
\end{equation}
where $\rho'=\sum_i(B_i\rho B_i)$ is the measured density matrix $\rho$ with the PVM associated with $\pi$. Also, from the second to the third line we used the fact that $[\tr(B_i\rho B_i)]^2=\tr(B_i\rho B_i)^2$.

\begin{figure*}
    \centering
    \includegraphics[width=\columnwidth]{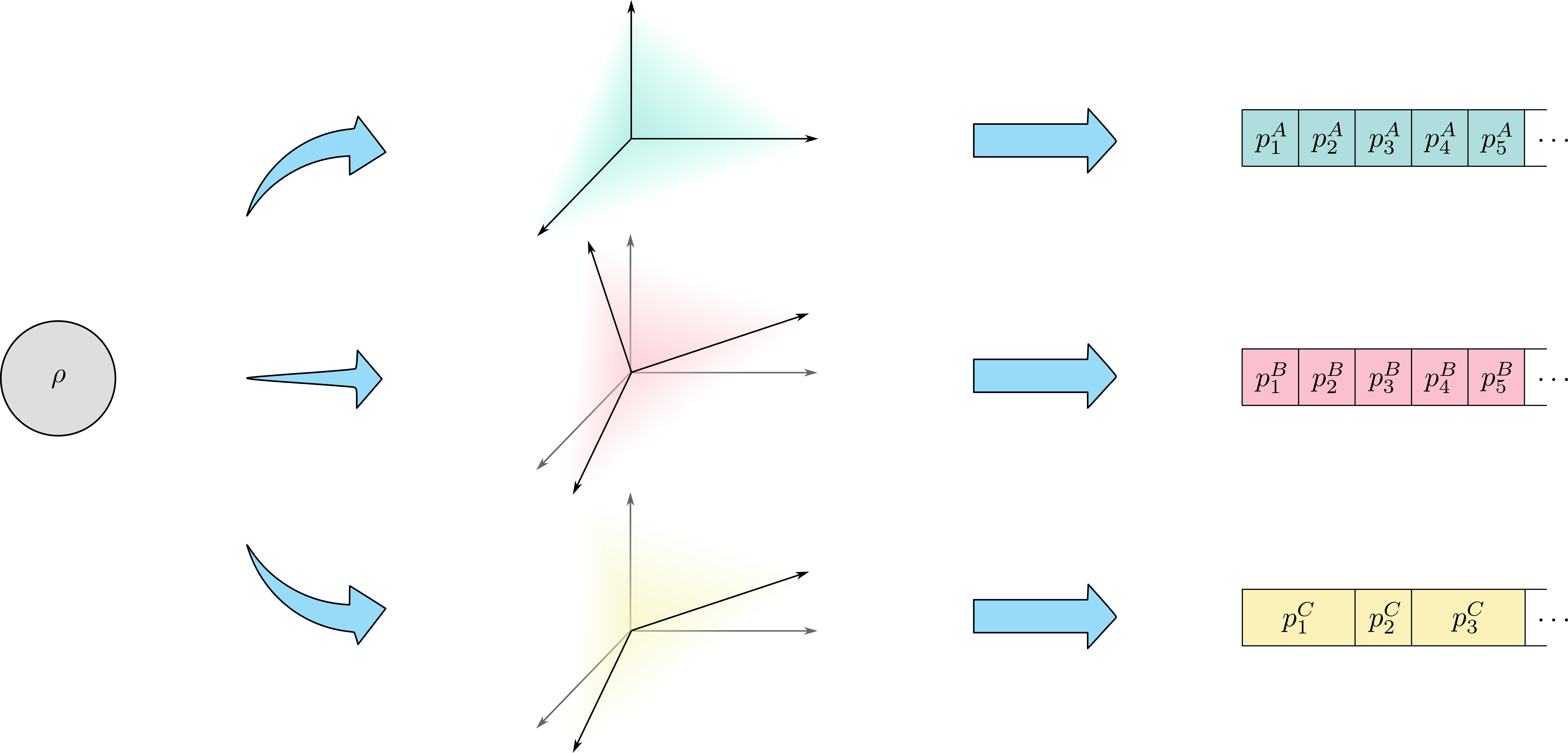}
    \caption{\textbf{Representation of the probability distribution used in the definition of quantum logical entropy.} Any state $\rho$, together with a partition characterized by a PVM, which is associated with an orthonormal basis in the corresponding Hilbert space, defines a probability distribution. This is the case even in case of coarse (i.e., degenerate) PVMs, as illustrated by the yellow partition. These distributions allow the introduction of (PVM-dependent) logical entropies for quantum systems.}
    \label{fig:qle}
\end{figure*}

The above already defines a logical entropy for the state $\rho$ when the partition $\pi$ (characterized by the PVM) is considered, as represented in Figure \ref{fig:qle}. Observe that this definition can be applied even in cases of coarse PVMs, i.e., PVMs with degeneracy. Moreover, differently from the von Neumann entropy, the logical entropy has a clear operational meaning.

The dependency on PVMs should not come as a surprise since, even classically, as already seen, the logical entropy, in general, depends on the partition. However, if we wish to have a PVM-independent quantum logical entropy, we can define
\begin{equation}
    L(\rho) \equiv \min_{\pi} \{L_\pi(\rho) \ | \ \pi \text{ is non-degenerate}\}.
    \label{eq:def-qle}
\end{equation}
While this is not the only possible definition, it is the one that is typically considered in the literature \cite{ellerman2018logical}. It is, also, in a certain sense, a ``natural'' choice. In fact, it turns out that the PVM that minimizes $L_\pi(\rho)$ is the one associated with a basis for which $\rho$ is diagonal, in which case $\rho=\rho'$. To see that, let $\rho = \sum_{ij} \rho_{ij} |i\rangle \langle j|$ be a representation of $\rho$ in the basis associated with the projectors $B_i$. In this same basis, $\rho' = \sum_i \rho_{ii} |i\rangle \langle i|$. Moreover,
\begin{equation}
    \tr\rho^2 = \sum_{ij} |\rho_{ij}|^2 = \tr{\rho'}^2 + \sum_i \sum_{j\neq i} |\rho_{ij}|^2,
    \label{eq:basis-comp}
\end{equation}
which implies that $L_\pi(\rho)=1-\tr{\rho}^2+\sum_i \sum_{j\neq i} |\rho_{ij}|^2$, which is minimized whenever $\rho$ is diagonal in the basis associated with the projectors $B_i$. Since this is the case for every non-degenerate PVM, we conclude that
\begin{equation}
    L(\rho) = \tr[\rho (I-\rho)]
    \label{eq:qle}
\end{equation}
and, as a result, expression \eqref{eq:basis-comp} can be rewritten as $L(\rho) = L(\rho') - \sum_i \sum_{j\neq i} |\rho_{ij}|^2$, which, in particular, leads to
\begin{equation}
    L(\rho)\leq L(\rho').
\end{equation}

Observe that the quantum logical entropy can be reduced to the complement of the \textit{purity} of the state $\rho$, which is given by $\gamma(\rho) = \tr\rho^2$. In fact, the expression for this entropy coincides with what is sometimes referred to as \textit{impurity} or \textit{mixedness} \cite{jaeger2007quantum}. Furthermore, similarly to the classical case, the logical entropy is also equivalent to the \textit{quantum Tsallis entropy of index two}, which, in turn, is equivalent to the so-called \textit{linear entropy} --- although its most appropriate name should be \textit{quadratic entropy}, as it was referred to in \cite{coles2011non}. While this seems to weaken the novelty in introducing the quantum logical entropy, observe that the connection with partitions, which is the starting point for that, may bring new insights and understanding to those already known quantities. Moreover, other definitions, like logical divergence and relative logical entropy that are presented next, are added on top of it, and they have the potential to help in the study of various scenarios in quantum information.

With that, we introduce the \textit{quantum logical divergence} as
\begin{equation}
    d(\rho||\sigma) = 2 \ \tr\rho(I-\sigma) - L(\rho) - L(\sigma).
    \label{eq:log-div}
\end{equation}
Note that there exists a relation between the logical divergence and \textit{quantum fidelity}, defined as
\begin{equation}
    F(\rho,\sigma) = \left(\tr\sqrt{\sqrt{\sigma}\rho\sqrt{\sigma}}\right)^2.
\end{equation}
In fact, definition \eqref{eq:log-div} can be rewritten as
\begin{equation}
    d(\rho||\sigma) = \gamma(\rho) + \gamma(\sigma) - 2 \tr(\rho\sigma).
\end{equation}
While the term $\tr(\rho\sigma)$ is not equivalent to $F(\rho,\sigma)$ in general, there exists an apparent similarity between them. This becomes more evident in cases where $\rho$ and $\sigma$ commute, for which $F(\rho,\sigma) = \left(\tr\sqrt{\rho\sigma}\right)^2$. Moreover, the two are equivalent if $\rho$ and $\sigma$ are pure states.

It should be noted that the logical divergence between a density matrix before and after a PVM measurement is simply
\begin{equation}
    d(\rho||\rho') = L(\rho') - L(\rho).
\end{equation}

Finally, we can also introduce the \textit{quantum relative logical entropy} as
\begin{equation}
    L(A/B) = L(\rho_{AB}) - L(I/d\otimes \rho_B),
\end{equation}
where $d$ is the dimension of $A$. Moreover, here and throughout the text, we denote $\rho_A=\tr_B\rho_{AB}$ and $\rho_B=\tr_A\rho_{AB}$.

Observe that the relative logical entropy can be reduced to a special case of the logical divergence. In fact, since
\begin{equation}
    tr[\rho_{AB}(I-I/d\otimes \rho_B)]= 1-\frac{1}{d} \tr\rho_B^2,
\end{equation}
and
\begin{equation}
    L(I/d\otimes \rho_B)= 1-\frac{1}{d} \tr\rho_B^2,
\end{equation}
it follows from the definition of both entropies that
\begin{equation}
    L(A/B) = -\frac{1}{4} d(\rho_{AB}||I/d\otimes \rho_B),
    \label{eq:rel-div}
\end{equation}
proving the relation between relative logical entropy and logical divergence. As a result, they share multiple properties.

In the following section, we state and prove various properties of quantum logical entropies.

\section{Properties of quantum logical entropies}

\noindent\textbf{Proposition 1 (Basic properties)}

\begin{itemize}
    \item[\textbf{(a)}] Logical entropy is non-negative and $L(\rho)=0$ for a pure state.
    \item[\textbf{(b)}] The maximal value of the logical entropy is $1-1/d$, where $d$ is the dimension of the Hilbert space. This value is the logical entropy of the maximally mixed state $I/d$.
    \item[\textbf{(c)}] Given a composite pure state $\rho_{AB}$ on the space $(A,B)$, it follows that $L(\rho_A)=L(\rho_B)$.
    \item[\textbf{(d)}] If $\rho_{AB} = \rho_A \otimes \rho_B$, then
    \begin{equation}
        L(\rho_A\otimes \rho_B) = L(\rho_A) + L(\rho_B)- L(\rho_A) \cdot L(\rho_B).
    \end{equation}
\end{itemize}

\noindent\textit{Proof}
\begin{itemize}
    \item[\textbf{(a)}] This follows from the definition since the logical entropy is a probability. However, another way to verify it is by observing that for every density matrix, $\tr\rho = 1$ and $\tr\rho^2 \le 1$, with equality if and only if $\rho$ is pure.

    \item[\textbf{(b)}] Let $\{\lambda_i\}$ be the set of eigenvalues of a density matrix $\rho$. Then, using the Cauchy-Schwarz inequality for two vectors $u$ and $v$ whose components are, respectively, $u_i=\lambda_i$ and $v_i=1/d$, it holds that $\tr\rho^2\geq 1/d$. Therefore, $L(\rho)\leq 1- 1/d$. Note also that $L(I/d) = 1-1/d$.

    \item[\textbf{(c)}] The result follows immediately from the Schmidt decomposition since $A$ and $B$ have the same orthonormal set of eigenvectors.

    \item[\textbf{(d)}] The result follows from writing $\rho_A$ and $\rho_B$ in their diagonal form and, then, using the identity
    \begin{equation}
        (1-x) + (1-y) - (1-x)(1-y) = 1-xy.
    \end{equation}
\end{itemize}
$\hfill\blacksquare$

Observe that part (d) of the previous proposition states, in particular, that the logical entropy of separable bipartite systems is subadditive, i.e.,
\begin{equation}
    L(\rho_{AB}) \leq L(\rho_A)+ L(\rho_B).
    \label{eq:subadditivity}
\end{equation}
The next proposition generalizes this result for generic bipartite systems.

\noindent\textbf{Proposition 2 (Subadditivity)} The logical entropy of a density matrix $\rho_{AB}$ is subadditive.

\noindent\textit{Proof}---This result was proved in 2007 by Audenaert for any quantum Tsallis entropy of index greater than one \cite{audenaert2007subadditivity}, which, in particular, includes the logical entropy. $\hfill\blacksquare$

Although the quantum logical entropy satisfies the subadditivity property, it should be noted that, differently from the von Neumann entropy, it does not satisfy the \textit{strong subadditivity}, i.e., in general, it does not hold that
\begin{equation}
    L(\rho_{ABC}) + L(\rho_B) \leq L(\rho_{AB})+ L(\rho_{BC}).
\end{equation}
However, the logical entropy satisfies a condition called \textit{firm subadditivity}, which is characterized as follows: Given a PVM defined by blocks $A_k$ in a subsystem $A$ of a bipartite state $\rho_{AB}$, the logical entropy is said to be firm subadditive if
\begin{equation}
    L(\rho_{AB}) \leq L(\rho_A) + \sum_k p_k L(\rho_B^{(k)}),
\end{equation}
where $p_k = \tr(A_k\rho_A)$ and $p_k\rho_B^{(k)} = \tr_A(A_k\rho_{AB})$.

\noindent\textbf{Proposition 3 (Firm subadditivity)} The logical entropy of a density matrix $\rho_{AB}$ is firm subadditive.

\noindent\textit{Proof}---This result was proved in 2011 by Coles in Theorem 5 of \cite{coles2011non}. $\hfill\blacksquare$

In our next result, we show that the logical entropy satisfies a triangle inequality.

\noindent\textbf{Proposition 4 (Triangle inequality)} For any density matrix $\rho_{AB}$, it holds that
\begin{equation}
    |L(\rho_A) - L(\rho_B)| \leq L(\rho_{AB}).
    \label{eq:triangular}
\end{equation}

\noindent\textit{Proof}---Let $R$ be a system such that $\rho_{ABR}$ is pure. From Proposition 2, we deduce that
\begin{equation}
    L(\rho_{BR}) \leq L(\rho_B) +L(\rho_R).
\end{equation}
Moreover, Proposition 1(c) implies that $L(\rho_R) = L(\rho_{AB})$ and $L(\rho_A) = L(\rho_{BR})$. As a result,
\begin{equation}
    L(\rho_A)- L(\rho_B) \leq L(\rho_{AB}).
\end{equation}
A similar reasoning leads to
\begin{equation}
    L(\rho_B)- L(\rho_A) \leq L(\rho_{AB}).
\end{equation}
Therefore, inequality \eqref{eq:triangular} holds. $\hfill\blacksquare$

As discussed earlier, by definition, a PVM cannot decrease the logical entropy of a system. Our next result shows that this is the case not only for PVMs but for any unital map $\Lambda_{U}$, which is a map that preserves the maximally mixed state, i.e., $\Lambda_{U} \left(I/d\right) = I/d$, where $d$ is the dimension of the Hilbert space. Note that for every positive operator-valued measure (POVM), which is given by a set of semi-definite matrices $\left\{E_{i}\right\} _{i=1}^n$ with $\sum_{i=1}^n E_{i} = I$, there are many corresponding implementations $M_{i}$ such that $E_{i} = M_i^\dag M_i$. $E_{i}$ are positive semidefinite and so $M_i = P_i U_i$ where $P_i$ is a unique Hermitian positive semidefinite (also denoted $P_i = \sqrt{E_i}$) and $U_i$ can be any unitary. Choosing $U_i = I$, we obtain a unital implementation $\Lambda_U\left(\rho\right) = \sum_{i=1}^n M_i \rho M_i^\dag$ of the POVM.

\noindent\textbf{Proposition 5 (Entropy after a unital map)} For any unital map $\Lambda_{U}$, it holds that
\begin{equation}
    L\left(\rho\right) \leq L\left(\rho'\right),
    \label{eq:unital-ineq}
\end{equation}
where $\rho'=\Lambda_{U}(\rho)$.

\noindent\textit{Proof}---Recall that given two vectors $x$ and $y$ of dimension $n$, $y$ is said to \textit{majorize} $x$, which is denoted by $x\prec y$, if $\sum_{i=1}^k x_i^{\downarrow} \leq \sum_{i=1}^k y_i^{\downarrow}$ for every $k\leq n$ and $\sum_{i=1}^{n} x_i = \sum_{i=1}^n y_i$, where $x^{\downarrow}$ is the vector $x$ reordered so that its components are in non-increasing order. Then, as stated in Lemma 3 of \cite{streltsov_maximal_2018} and proven in \cite{uhlmann1970shannon, nielsen2002introduction}, given two density matrices $\rho$ and $\sigma$ of equal dimension, $\rho$ can be converted into $\sigma$ via some unital map $\Lambda$ if and only if the vector of eigenvalues of $\rho$ majorizes the vector of eigenvalues of $\sigma$. In our case of interest, if $x$ is the vector of eigenvalues of $\rho'$ and $y$ is the vector of eigenvalues of $\rho$, $x\prec y$.

Moreover, as shown in Proposition 12.11 of \cite{nielsen2000quantum}, $x\prec y$ if and only if $x = \sum_{j=1}^m p_j P_j y$ for some probability distribution $p_j$ and permutation matrices $P_j$. Now, letting $P_{j}$ be the permutation matrix associated with the permutation $\sigma_{j}$, we can write $x_i = \sum_{j=1}^m p_j y_{\sigma_{j}(i)}$. As a result, $x^2 = \sum_{i=1}^d x_i^2$ gives
\begin{equation}
    \begin{aligned}
        x^2 &= \sum_{i=1}^d \left(\sum_{j=1}^m p_j y_{\sigma_{j}(i)}\right)^2 \\
            &= \sum_{i=1}^d \sum_{j=1}^m p_j^2 y_i^2 + \sum_{i=1}^d \sum_{j=1}^m \sum_{k\neq j} p_j p_k y_{\sigma_j(i)} y_{\sigma_k(i)}.
    \end{aligned}
\end{equation}
Furthermore, since $\sum_{j=1}^m p_j = 1$,
\begin{equation}
    \begin{aligned}
        y^2 &= \sum_{i=1}^{d}\left(\sum_{j=1}^m p_j\right)^2 y_i^2 \\
            &= \sum_{i=1}^d \sum_{j=1}^m p_j^2 y_i^2 + \sum_{i=1}^d \sum_{j=1}^m \sum_{k\neq j} p_j p_k y_i^2.
    \end{aligned}
\end{equation}
Therefore,
\begin{equation}
    \begin{aligned}
        y^2-x^2 &= \sum_{i=1}^d \sum_{j=1}^m \sum_{k\neq j} p_j p_k \left(y_i^2 - y_{\sigma_j(i)} y_{\sigma_k(i)}\right) \\
        &= \sum_{i=1}^d \sum_{j=1}^m \sum_{k\neq j} p_j p_k \left(\frac{1}{2} y_{\sigma_{j}(i)}^2 + \frac{1}{2} y_{\sigma_{k}(i)}^2 - y_{\sigma_{j}(i)} y_{\sigma_{k}(i)} \right) \\
        &= \frac{1}{2} \sum_{i=1}^d \sum_{j=1}^m \sum_{k\neq j} p_j p_k \left(y_{\sigma_{j}(i)} - y_{\sigma_{k}(i)}\right)^2 \\
        &\geq 0.
    \end{aligned}
\end{equation}
Since $x$ and $y$ are, respectively, the vectors of the eigenvalues of $\rho'$ and $\rho$,
\begin{equation}
    \tr\left(\rho'\right)^{2} \leq \tr\rho^2,
\end{equation}
which implies that inequality \eqref{eq:unital-ineq} holds. $\hfill\blacksquare$

Suppose a system of interest interacts unitarily with other systems whose individual states are not under experimental control, e.g., the environment. Such interactions can be effectively represented by unital maps acting on the system of interest, i.e., a system of interest that starts in the state $\rho_S$ can be repersented by $\rho'_S = \sum_i E_i \rho_S E_i^\dag$ after the interaction. Then, from the previous proposition, the system of interest's logical entropy increases. In particular, the lower bound for the logical entropy should no longer be null. This lower bound as well as an upper bound in case the joint system is pure is the concern of the next result.

\noindent\textbf{Proposition 6 (Entropy after unitary interaction)} Assume a system $S$ interacts with a system $R$ through a unitary $U$. Also, let the joint system after the interaction be $\rho'_{SR} = U\rho_{SR}U^\dag$ and define $B_{ij} = \langle i| \rho'_{SR} |j\rangle$, where $\rho_{SR}$ is the state of the joint system before the interaction and $\{|i\rangle\}$ is an orthonormal basis of system $R$. Then,
\begin{equation}
    L(\rho'_S) \geq 2 \sum_i \sum_{j<i} \left\{\tr(B_{ij} B_{ij}^\dag) - \text{Re}[\tr(B_{ii}B_{jj})]\right\},
    \label{eq:ineq}
\end{equation}
where $\rho'_S = \tr_R \rho'_{SR}$. Furthermore, if $\rho_{SR}$ is a pure state,
\begin{equation}
    L(\rho'_S) \leq 2 \sum_i \sum_{j<i} \tr(B_{ij} B_{ij}^\dag).
    \label{eq:ineq2}
\end{equation}

\noindent\textit{Proof}---Observe that $\rho'_S = \sum_i B_{ii}$ and hence ${\rho'_S}^2 = \sum_{ij} B_{ii}B_{jj}$. As a result,
\begin{equation}
    \begin{aligned}
        L(\rho'_S) &= 1 - \sum_i \tr(B_{ii}^2) - \sum_i \sum_{j\neq i} \tr(B_{ii}B_{jj}) \\
                   &= 1 - \sum_i \tr(B_{ii}^2) - 2 \sum_i \sum_{j<i} \text{Re}[\tr(B_{ii}B_{jj})].
    \end{aligned}
    \label{eq:ls}
\end{equation}
Moreover, $\rho'_{SR} = \sum_{ij} B_{ij} \otimes |i\rangle\langle j|$ and ${\rho'_{SR}}^2 = \sum_{ijk} B_{ik} B_{kj} \otimes |i\rangle\langle j|$. Then, since $\tr{\rho'_{SR}}^2 \leq 1$,
\begin{equation}
    1 - \sum_i \tr(B_{ii}^2) \geq \sum_i \sum_{j\neq i} \tr(B_{ij} B_{ji}).
    \label{eq:lsr}
\end{equation}
Finally, combining expressions \eqref{eq:ls} and \eqref{eq:lsr}, we obtain inequality \eqref{eq:ineq}, which completes the first part of the proof.

Now, if $\rho_{SR}$ is a pure state, which implies that $\rho'_{SR}$ is also pure, $\tr{\rho'_{SR}}^2=1$ and, as a result, expression \eqref{eq:lsr} becomes an equality. The left-hand side of the latter is greater or equals to $L(\rho'_S)$, as can be directly seen from \eqref{eq:ls}. Therefore, we are lead to inequality \eqref{eq:ineq2} holds, completing the proof. $\hfill\blacksquare$

\noindent\textbf{Proposition 7 (Entropy of classical mixtures of quantum systems)} Let $\rho = \sum_i p_i \rho_i$, then
\begin{equation}
    L(\rho) \leq L(\{p_i\}) + \sum_k p_k^2 L(\rho_k),
\end{equation}
where $L(\{p_i\})$ is defined in \eqref{eq:classical-le}. Moreover, the equality holds whenever the matrices $\rho_i$ have orthogonal support.

\noindent\textit{Proof}---To start, assume the states $\rho_i$ are pure, i.e., $\rho_i = \ket{\psi_i} \bra{\psi_i}$. Then, $\rho = \sum_i p_i \ket{\psi_i} \bra{\psi_i}$. Introducing an auxiliary system $R$ that purifies $\rho$, the state of the joint system can be written as $\ket{\Psi_{SR}} = \sum_i \sqrt{p_i} \ket{\psi_i} \ket{i}$, where $\{\ket{i}\}$ is an orthonormal basis of system $R$. Observe that $\rho=\rho_S$ and
\begin{equation}
    \rho_R = \sum_{ij} \sum_\ell \left(\bra{\ell} \psi_i \rangle \bra{\psi_j} \ell\rangle \right) \ket{i}\bra{j} \sqrt{p_i} \sqrt{p_j},
\end{equation}
where $\{\ket{\ell}\}$ is an orthonormal basis of $S$. Also, by Proposition 1(c), $L(\rho_S) = L(\rho_R)$.

If $\rho_R$ is measured with a PVM characterized by $P_i = \ket{i}\bra{i}$, it holds that
\begin{equation}
    \rho'_R = \sum_i P_i \rho_R P_i = \sum p_i \ket{i}\bra{i}.
\end{equation}
This implies that $L(\rho'_R)= L(\{p_i\})$. Since PVMs increase the logical entropy,
\begin{equation}
    L(\{p_i\}) = L(\rho'_R) \geq L(\rho_R) =  L(\rho_S).
\end{equation}
Since the logical entropy of pure states vanish, the above expression proves the proposition for cases where the matrices $\rho_i$ are pure states.

For the general case, let $\rho_i = \sum_{ij} \lambda_{ij} \ket{\lambda_{ij}} \bra{\lambda_{ij}}$, where the states $\ket{\lambda}_{ij}$ are orthonormal vectors in the subspace associated with $\rho_i$. Hence, $\rho$ can be written as the sum of pure states $\rho = \sum_{ij} p_i \lambda_{ij} \ket{\lambda_{ij}} \bra{\lambda_{ij}}$ (not necessarily all orthogonal). Now, using the result for pure states we just obtained, it hols that
\begin{equation}
    \begin{aligned}
        L(\rho) &\leq L(\{p_i \lambda_{ij}\}) \\
        &= 1-\sum_{ij} p_i^2 \lambda_{ij}^2 \\
        &= 1-\sum_{i} p_i^2 + \sum_k p_k^2 (1-\sum_j \lambda_{kj}^2) \\
        &= L(\{p_i\}) + \sum_k p_k^2 L(\rho_k),
    \end{aligned}
\end{equation}
as we wanted to show.

Finally, it follows from direct computation that the equality holds whenever the matrices $\rho_i$ have orthogonal support. In fact, in this case, $\rho^2 = \sum_i p_i^2 \rho_i^2$ and, as a result,
\begin{equation}
    \begin{aligned}
        L(\rho) &= 1- \sum_i p_i^2 \tr\rho_i^2 \\
        &= 1- \sum_i p_i^2 + \sum_k p_k^2 (1-\tr\rho_k^2),
    \end{aligned}
\end{equation}
which concludes the proof. $\hfill\blacksquare$

In the next result, we prove the non-negativity of the logical divergence.

\noindent\textbf{Proposition 8 (Klein's inequality)} The logical divergence is always non-negative, i.e.,
\begin{equation}
    d(\rho||\sigma)\geq 0
\end{equation}
for every pair of density matrices $\rho$ and $\sigma$, with equality holding if and only if $\rho=\sigma$.

\noindent\textit{Proof}---It follows by direct computation that
\begin{equation}
    d(\rho||\sigma) = \tr(\rho-\sigma)^2.
    \label{eq:norm}
\end{equation}
Then, recall that for any Hermitian matrix $A$, $\tr(A^2) \geq 0$, with $\tr(A^2)=0$ if and only if $A=0$. $\hfill\blacksquare$

It should be noted that \eqref{eq:norm} defines the logical divergence as the square of the Hilbert-Schmidt norm of the difference of the two density matrices under consideration. Moreover, it can be shown that this norm coincides with a natural definition of the Hamming distance between density matrices \cite{ellerman2018logical}.

In the next three propositions, we study the concavity of logical entropies.

\noindent\textbf{Proposition 9 (Concavity of logical entropy)} Let $\rho = \sum p_i \rho_i$, where $\rho_i$ is a density matrix for each $i$ and $0\le p_i \le 1$, such that $\sum_i p_i = 1$. Moreover, set $\overline{L(\rho)} = \sum p_i L(\rho_i)$.
\begin{itemize}
    \item[\textbf{(a)}] If $\rho_i$ have orthogonal support, then
    \begin{equation}
        \overline{L(\rho)}< L(\rho).
        \label{eq:concavity1}
    \end{equation}
    \item[\textbf{(b)}] In general,
    \begin{equation}
        \overline{L(\rho)}- L(\{p_i\}) < L(\rho) < \overline{L(\rho)}+ L(\{p_i\}),
        \label{eq:concavity2}
    \end{equation}
    where $L(\{p_i\})$ is the classical logical entropy of the distribution $\{p_i\}_i$. In other words, $L(\rho)$ is in the $L(\{p_i\})$ neighborhood of $\overline{L(\rho)}$.
\end{itemize}

\noindent\textit{Proof}
\begin{itemize}
    \item[\textbf{(a)}] We will demonstrate the argument on two density matrices $\rho_1$ and $\rho_2$ having an orthogonal support. Set $\rho_1 = \sum_i p_i \ket{i}\bra{i} $, $\rho_2 = \sum_j q_j \ket{j}\bra{j}$, where $\ket{i}$ and $\ket{j}$ are two bases with orthogonal support, also set $\rho = \lambda \rho_1 + (1-\lambda) \rho_2$, where $0<\lambda<1$. Now,
    \begin{equation}
        \begin{aligned}
            \overline{L(\rho)} &= \lambda L(\rho_1) + (1-\lambda) L(\rho_2) \\
            &= \lambda  ( 1- \sum_i p_i^2) + (1-\lambda) (1- \sum_j q_j^2) \\
            &= 1-\lambda \sum_i p_i^2 - (1-\lambda)\sum_j q_j^2.
        \end{aligned}
    \end{equation}
    However,
    \begin{equation}
        L(\rho) = 1-\lambda^2 \sum_i p_i^2 - (1-\lambda)^2\sum_j q_j^2.
    \end{equation}
    Therefore, inequality \eqref{eq:concavity1} holds.

    \item[\textbf{(b)}] Consider $\rho_{AB} = \sum_i p_i \rho_i\otimes \ket{i} \bra{i}$, so $\rho_{AB}$ is a sum of densities with an orthogonal support. From the result in part (a) and Proposition 2, it follows that
    \begin{equation}
        \overline{L(\rho)} \leq L (\rho_{AB}) \leq L(\rho_A) + L(\rho_B).
    \end{equation}
    However, $\rho_A = \rho$ and $L(\rho_B)= L(\{p_i\})$, which leads to
    \begin{equation}
        \overline{L(\rho)} - L(\{p_i\}) \leq L(\rho),
    \end{equation}
    proving the first part of inequality \eqref{eq:concavity2}. To conclude the proof, we first show that $L(\rho)\leq \overline{L(\rho)} + L(\{p_i\})$ for the case where the matrices $\rho_i$ are pure states, i.e., $\rho_i = \ket{\psi_i} \bra{\psi_i}$. For that, consider the following purification of the system:
    \begin{equation}
        \ket{\eta} = \sum_i \sqrt{p_i} \ket{\psi_i} \otimes \ket{i},
    \end{equation}
    where the vectors $\ket{i}$ are orthonormal in some system $B$. Moreover, denote $\tilde{\rho} = \ket{\eta}\bra{\eta}$. It is clear, then, that $\rho = \tilde{\rho}_A = \sum_i p_i \ket{\psi_i} \bra{\psi_i}$ and
        \begin{equation}
            \tilde{\rho}_B = \sum_{ij} \sqrt{p_i p_j} \bra{\psi_j} \psi_i\rangle \ket{i}\bra{j}.
        \end{equation}
    Note that the vectors $\ket{\psi_i}$ are not necessarily orthogonal. Measuring system $B$ with the operators $P_i = \ket{i}\bra{i}$, we obtain $\tilde{\rho}_B' = \sum_i p_i \ket{i} \bra{i}$. By Proposition 5, $L(\tilde{\rho}_B') = L(\{p_i\}) \geq L(\tilde{\rho}_B) = L(\rho)$, where we used Proposition 1(c) in the last step. Therefore, for $\rho$ which is a sum of pure states, we have $L(\rho)\leq L(\{p_i\})$, which is consistent with \eqref{eq:concavity2} since, in this case, $\overline{L(\rho)}=0$.

    Consider now the general case where $\rho = \sum_i p_i \rho_i$ with
    \begin{equation}
        \rho_i = \sum_j p_i^j \ket{e_i^j} \bra{e_i^j}.
    \end{equation}
    Here, $\{\ket{e_i^j}\}_j$ are orthonormal vectors for each $i$. Hence,
    \begin{equation}
        \rho = \sum_{ij} p_i p_i^j \ket{e_i^j} \bra{e_i^j},
    \end{equation}
    where $\ket{e_i^j} \bra{e_i^j}$ are pure states for every $i$ and $j$. We can use the previous result for pure states to conclude that
    \begin{equation}
        L(\rho) \leq L(\{p_i p_i^j\}) = \sum_{ij} p_i p_i^j (1-p_i p_i^j).
    \end{equation}
    Using the fact that, for $x_i,x_j\in[0,1]$,
    \begin{equation}
        1-x_i x_j \leq (1-x_i) + (1-x_j),
    \end{equation}
    we are lead to
    \begin{equation}
        \begin{aligned}
            L(\rho) &\leq \sum_{ij} p_i p_i^j (1-p_i)+ \sum_{ij} p_i p_i^j (1-p_i^j) \\
            &= L(\{p_i\}) + \sum_i p_i L(\rho_i),
        \end{aligned}
    \end{equation}
    where we used the orthogonality of the set of vectors $\{\ket{e_i^j}\}_j$ for each $i$. $\hfill\blacksquare$
\end{itemize}

\noindent\textbf{Proposition 10 (Joint convexity of logical divergence)} The logical divergence $d(\rho||\sigma)$ is jointly convex.

\noindent\textit{Proof}---We start by defining $\rho = \lambda \rho_1 + (1-\lambda) \rho_2$ and $\sigma = \lambda \sigma_1 + (1-\lambda) \sigma_2$. Then, it follows by direct computation that
\begin{equation}
    \begin{aligned}
        d(\rho||\sigma) &= \tr(\rho - \sigma)^2 \\
        &= \tr[\lambda(\rho_1-\sigma_1) + (1-\lambda)(\rho_2-\sigma_2)]^2 \\
        &\leq \lambda \tr(\rho_1-\sigma_1)^2 + (1-\lambda) \tr(\rho_2-\sigma_2)^2 \\
        &= \lambda d(\rho_1||\sigma_1) +(1-\lambda) d(\rho_2||\sigma_2),
    \end{aligned}
\end{equation}
where the inequality is due to the convexity of $\tr(x^2)$. $\hfill\blacksquare$

\noindent\textbf{Proposition 11 (Concavity of relative entropy)} The relative logical entropy $L(A/B)$ is a concave function of $\rho_{AB}$.

\noindent\textit{Proof}---It follows direct from the relation given by \eqref{eq:rel-div} and Proposition 10. $\hfill\blacksquare$

The next proposition states the fact that the divergence behaves as a metric. Tracing out a subspace can only reduce its value.

\noindent\textbf{Proposition 12 (Monotonicity of logical divergence)} Let $\rho_{AB}$ and $\sigma_{AB}$ be two density matrices, then
\begin{equation}
    d(\rho_A\otimes I/b||\sigma_A\otimes I/b) \leq d(\rho_{AB}||\sigma_{AB}),
\end{equation}
where $b$ is the dimension of $B$.

\noindent\textit{Proof}---As can be seen in Chapter 11 of \cite{nielsen2000quantum}, there exist a set of unitary matrices $U_j$ over $B$ and a probability distribution $p_j$ such that
\begin{equation}
    \rho_A \otimes I/b = \sum_j p_j U_j \rho_{AB} U_j^\dag.
\end{equation}
for every $\rho_{AB}$. Then, because the logical divergence is jointly convex on both entries, we can write
\begin{equation}
d(\rho_A\otimes I/b||\sigma_A\otimes I/b) \leq \sum_j p_j  d(U_j \rho_{AB} U_j^\dag||U_j \sigma_{AB} U_j^\dag).
\end{equation}
Finally, since the divergence is invariant under unitary transformations, the above sum gives $d(\rho_{AB}||\sigma_{AB})$. $\hfill\blacksquare$

\section{Quantum logical entropy of post-selected systems}

We shall now propose extensions of the notion of quantum logical entropy to the class of pre- and post-selected quantum systems. Like before, we use the fact that every PVM defines a partition $\pi$ in a Hilbert space given by a basis $\{|b_i\rangle\}$ associated with it since every PVM is characterized by parts $B_i=|b_i\rangle\langle b_i|$ such that $\sum_i B_i = I$. Suppose that a system is prepared in a state $|\psi\rangle$ and a PVM is performed. Moreover, assume that it is known that the system is later found in a state $|\phi\rangle$ non-orthogonal to $|\psi\rangle$. In scenarios like that, with a pre-selected $|\psi\rangle$ and a post-selected $|\phi\rangle$, the system can be represented by the generalized density matrix
\begin{equation}
    \rho_{\psi|\phi} = \frac{|\psi\rangle\langle\phi|}{\langle\phi|\psi\rangle}.
\end{equation}
The probability of measuring a result associated with $B_i$ can be computed as $|\tr(B_i\rho_{\psi|\phi})|^2 = \tr(B_i\rho_{\psi|\phi}B_i\rho_{\psi|\phi}^\dag)$. This result is known as the ABL rule \cite{aharonov1964time}. Then, in analogy with the definition of logical entropy for classical and non-post-selected quantum systems, we may define
\begin{equation}
    L_\pi(\rho_{\psi|\phi}) \equiv \sum_i |\tr(B_i\rho_{\psi|\phi})|^2 |1 - \tr(B_i \rho_{\psi|\phi})|^2.
\end{equation}
Since $|1 - \tr(B_i \rho_{\psi|\phi})|^2 = 1 - \tr(B_i \rho_{\psi|\phi} B_i) - \tr(B_i \rho_{\psi|\phi}^\dag B_i) + \tr(B_i \rho_{\psi|\phi} B_i \rho_{\psi|\phi}^\dag B_i)$, it holds that
\begin{equation}
        \begin{aligned}
        L_\pi(\rho_{\psi|\phi}) &= \sum_i \tr|B_i\rho_{\psi|\phi}B_i (I - B_i \rho_{\psi|\phi} B_i)|^2 \\
        &= |\tr[\rho'_{\psi|\phi} (I-\rho'_{\psi|\phi})]|^2,
    \end{aligned}
\end{equation}
where $\rho'_{\psi|\phi} = \sum_i B_i\rho_{\psi|\phi} B_i$.

Differently from the case without post-selection, where a PVM-independent logical entropy could be defined by minimizing over non-degenerate PVMs, this is not possible here. The reason is that we are interested in cases where the pre- and post-selections are pure states. Then, although the minimum $L_\pi(\rho_{\psi|\phi})$ corresponds to the case where $\tr(\rho'_{\psi|\phi}) = \tr(\rho_{\psi|\phi})$, it also corresponds to the null function. This is the case because in such scenarios the intermediate PVM is associated with an orthonormal basis that contains either $|\psi\rangle$ or $|\phi\rangle$. As a result, these PVMs allow us to see $\rho_{\psi|\phi}$ as a partition with a single part.

Observe that, by construction, the logical entropy for post-selected systems is always positive. This fact contrasts with a generalization of the von Neumann entropy proposed in \cite{salek2014negative}, which can assume negative values for some post-selections.

Another approach for this problem employs directly the notion of \textit{weak values} \cite{aharonov1988result}, which has been shown to be very constructive in the study of pre- and post-selected systems \cite{dixon09, jacobs09, turner11, aharonov14, dressel2014colloquium, alves2015weak, cortez2017rapid, kim18, naghiloo18, Hu18, pfender19, cujia19, paiva2021aharonov}. This approach consists of replacing the intermediate PVM by the inference of the weak values of orthogonal projectors associated with a specific PVM. Given a pre-selection $|\psi\rangle$, a post-selection $|\phi\rangle$, and a PVM characterized by parts (projectors) $B_i$, the weak value associated with each operator $B_i$ is defined as $\tr(B_i\rho_{\psi|\phi})$. Since $\sum_i \tr(B_i\rho_{\psi|\phi})$, the weak values of this set of projectors can be seen as a quasi-probability distribution. It is possible, then, to use these quasi-probabilities to construct a weak logical entropy for post-selected quantum systems, which gives
\begin{equation}
    \begin{aligned}
        L_\pi^w(\rho_{\psi|\phi}) &\equiv \sum_i \tr(B_i\rho_{\psi|\phi}) \tr[(I-B_i)\rho_{\psi|\phi}] \\
        &= \sum_i \tr(B_i\rho_{\psi|\phi}) [1 - \tr(B_i\rho_{\psi|\phi})] \\
        &= \tr[\rho'_{\psi|\phi} (I-\rho'_{\psi|\phi})].
    \end{aligned}
\end{equation}
Here, again, a PVM-independent logical entropy cannot be defined. However, just like the extension of the von Neumman entropy to post-selected systems \cite{salek2014negative}, the weak logical entropy can be negative. Even further, it may be complex-valued.

While the operational meaning of $L_\pi(\rho_{\psi|\phi})$ is clear, it is not straightforward to interpret $L_\pi^w(\rho_{\psi|\phi})$. One of the reasons is that weak values are associated with weak measurements, and weak measurements do not divide the Hilbert space into disjoints parts. This fact in itself can be also seen as the reason why we end up with a quasiprobability, and not a probability. However, consider the following game: A system is prepared in a state $|\psi\rangle$. Later, a weak measurement of one of the blocks $B_i$ of a PVM characterized by them. The system is then post-selected in the state $|\phi\rangle$. If this process is repeated and the PVM is composed by $n$ blocks, then the probability that the weak measurement of a different block is performed is simply $(n-1)/n$. However, suppose that, instead, we wanted to refer to the obtained weak value as the quasiprobability associated with each block. In this case, keeping an algebraic analogy to the classical logical entropy defined in \eqref{eq:cle-part} and the PVM-dependent quantum logical entropy defined in \eqref{eq:qle-part}, we obtain the weak logical entropy $L_\pi^w(\rho_{\psi|\phi})$.

Interestingly, even though the logical and the weak logical entropies of post-selected systems are associated with different operational meanings, it should be noted that
\begin{equation}
    L_\pi(\rho_{\psi|\phi}) = |L_\pi^w(\rho_{\psi|\phi})|^2.
    \label{eq:relation}
\end{equation}
Thus, in this sense, the weak logical entropy is mathematically more fundamental than the logical entropy of post-selected systems first introduced in this section.

\section{Discussion}

In this work we presented a new path to the introduction of quantum logical entropy and proved various properties of it. As discussed above, the PVM-independent definition of this entropy is equivalent to other already known quantities. However, the perspective of partitions that comes with its introduction may bring new insights into its use in the study of quantum systems within different scenarios. Furthermore, it may lead to the discovery of new implications for the logical structure, uncertainty and information processing associated with quantum systems.

In fact, the properties of logical entropies proved here may be relevant to quantum information science and technology as well as other areas of quantum mechanics, as already shown, e.g., in \cite{zurek1993coherent, buscemi2007linear, tamir2015holevo}. In particular, a special case of Proposition 6 concerns quantum systems in the presence of noise. In this case, the logical entropy of the system of interest after its interaction with the noise is bounded by correlations between this system and its environment. This scenario, including an example of amplitude damping, was considered in \cite{tamir2017tsallis}. Moreover, since the logical entropy is a measure of distinction, it seems natural to use it in the investigation of channel capacity in terms of quantum dits. We expect that the language of quantum dits can simplify the proofs of channel properties. It would also be interesting to examine the use of logical entropy in the context of entanglement quantification in discrete/continuous systems.

Notably, while some of the properties of logical entropies are similar to the ones held by the more widely studied von Neumann entropy, many of them also evidence major differences. It stands out, for instance, that the logical entropy does not fulfill the strong subadditivity property. Even though it still satisfies subadditive properties, as stated in Propositions 2 and 3, the ``breakdown'' of strong additivity implies an unmistakable departure from the von Neumann entropy. The lack of this property might at first appear somewhat troublesome. However, it may also place the quantum logical entropy as a fundamental player in various special scenarios, e.g., when discussing the breakdown of subadditivity in black holes \cite{almheiri2013black}.

Finally, we briefly extended the notion of logical entropy to post-selected quantum systems in two ways. In the first, there exists a clear operational meaning associated with the logical entropy. In the second (weak logical entropy), there exists only a loose connection with experimental procedures associated with weak measurements. It would be important to study better the significance of the latter. Furthermore, properties of these logical entropies need yet to be investigated. Also, a relation between both definitions can be established as seen in \eqref{eq:relation}, where the square of the magnitude of the weak logical entropy gives the logical entropy of post-selected systems. This raises a question about the phase of the weak logical entropy. While we could not find a direct meaning to it, it seems to us that this problem deserves further analysis.

\section*{Acknowledgements}
We thank D. Ellerman and Y. Neuman for many discussions on the topic as well as their helpful comments. We are also thankful to T. Landsberger and J. Kupferman for helpful feedback regarding a previous version of this manuscript. E.C. was supported by Grant No. FQXi-RFP-CPW-2006 from the Foundational Questions Institute and Fetzer Franklin Fund, a donor advised fund of Silicon Valley Community Foundation, the Israel Innovation Authority under Projects No. 70002 and No. 73795, the Quantum Science and Technology Program of the Israeli Council of Higher Education, and the Pazy Foundation.

\bibliography{citations}

\end{document}